\documentclass[letterpaper, conference]{IEEEtran}
\IEEEoverridecommandlockouts
\usepackage{authblk}
\usepackage{cite}
\usepackage{amsmath,amssymb,amsfonts}
\usepackage{algorithmic}
\usepackage{graphicx}
\usepackage{textcomp}
\usepackage{xcolor}
\usepackage{subcaption}
\usepackage[hidelinks]{hyperref}   
\usepackage{pgfplots}
\usepackage[inline]{enumitem}
\usepackage{tikz}    
\usetikzlibrary{shapes,calc,backgrounds,positioning,arrows.meta,patterns}
  
\usepackage{flushend}
  
\definecolor{myYellow}{RGB}{253, 231, 37}  
\definecolor{myLime}{RGB}{173,220,48}  
\definecolor{myGreen}{RGB}{94,201, 98}  
\definecolor{myGreenish}{RGB}{40,174,128}  
\definecolor{myTeal}{RGB}{33, 145, 140}  
\definecolor{myOcean}{RGB}{44,114,142}  
\definecolor{myBlue}{RGB}{59, 82, 139}  
\definecolor{myPurple}{RGB}{71,45,123}  
\definecolor{myEggplant}{RGB}{68, 1, 84}  
\pgfplotsset{compat=1.18}  
\usepgfplotslibrary{colormaps}  
\pgfplotsset{/pgfplots/bar cycle list/.style={/pgfplots/cycle list={%
    {myTeal,fill=myTeal!80!white,mark=none,postaction={pattern=north west lines, pattern color=myTeal!70!black}},%
    {myEggplant!60!black,fill=myEggplant!80!white,mark=none,postaction={pattern=north east lines, pattern color=myEggplant!70!black}},
    {myYellow,fill=myYellow!80!white,mark=none,postaction={pattern=crosshatch,pattern color=myYellow!70!black}},
    {myGreen,fill=myGreen!80!white,mark=none,postaction={pattern=vertical,pattern color=myGreen!70!black}},  
    {myblue,fill=myBlue!80!white,mark=none,postaction={pattern=grid,pattern color=myBlue!70!black}},  
},},}                            
  
\pgfplotsset{
    ourybarstyle/.style={
        axis x line*=bottom,  
        axis y line*=left,           
        ymin=0,  
        xtick=data,  
        ylabel style={at={(yticklabel* cs:1)}, anchor=south east, rotate=-90},  
        nodes near coords,  
        nodes near coords align={vertical},  
    },  
    ourbarstylenonums/.style={
        axis x line*=bottom,  
        axis y line*=left,           
        ymin=0,  
        xtick=data,  
        ylabel style={at={(yticklabel* cs:1)}, anchor=south east, rotate=-90},  
    },
    ourlinestyle/.style={
        ymin=0,  
    },  
}  

\def\BibTeX{{\rm B\kern-.05em{\sc i\kern-.025em b}\kern-.08em
    T\kern-.1667em\lower.7ex\hbox{E}\kern-.125emX}}

\newcommand{\CLASSINPUTbottomtextmargin}{0.99in}

\newcommand{\OSname}{Ariel~OS}
\newcommand{\espcthree}{ESP32\nobreakdash-C3}
\newcommand{\espsthree}{ESP32\nobreakdash-S3}

\newcommand\noteEB[1]{\textcolor{red}{EB: #1}}
\newcommand\noteEF[1]{\textcolor{blue}{EF: #1}}
\newcommand\noteKS[1]{\textcolor{brown}{KS: #1}}

\begin{document}

\title{

 \footnotesize
 \framebox[1.01\width]{\parbox{\dimexpr\linewidth-2\fboxsep-2\fboxrule}{If you cite this paper, please use: \emph{E. Frank et al.} Ariel OS: An Embedded Rust Operating System for Networked Sensors \& Multi-Core Microcontrollers. \emph{In Proc. of the 21st IEEE International Conference on Distributed Computing in Smart Systems and the Internet of Things} (DCOSS-IoT), June 2025.}}
  \ \\ \ \\ \ \\
 \Huge Ariel OS: An Embedded Rust Operating System for Networked Sensors \& Multi-Core Microcontrollers
}

\author[2]{E. Frank}
\author[1,2]{K. Schleiser}
\author[1]{R. Fouquet}
\author[2]{K. Zandberg}
\author[1,3]{C. Ams\"{u}ss}
\author[1,2,3]{E. Baccelli}
\affil[1]{Inria, France}
\affil[2]{Freie Universit\"{a}t Berlin, Germany}
\affil[3]{Einstein Center Digital Future, Germany}



\maketitle
\IEEEpeerreviewmaketitle

\begin{abstract}
Large swaths of low-level system software building blocks originally implemented in C/C++ are currently being swapped for equivalent rewrites in Rust, a relatively more secure and dependable programming language. 
So far, however, no embedded OS in Rust 
supports multicore preemptive scheduling on microcontrollers.
In this paper, we thus fill this gap with a new operating system:  \OSname{}. 
We describe its design, we provide the source code of its implementation, and we perform micro-benchmarks on the main 32-bit microcontroller architectures: ARM Cortex-M, RISC-V and Espressif Xtensa. 
We show how our scheduler takes advantage of several cores, while incurring only small overhead on single-core hardware. 
As such, \OSname{} provides a convenient embedded software platform for small networked devices, for both research and industry practitioners.
\end{abstract}


\section{Introduction}

We increasingly depend on cyberphysical systems and distributed computing systems which encompass 
not only machines in the microprocessor segment, but also much more resource-constrained devices such as sensors based on microcontroller units (MCU). The latter, as characterized in RFC7228~\cite{rfc7228}, achieve ultra-low power consumption and ultra-low price tags, but provide much less memory (in the \emph{kilobyte} range) and much weaker processing power (CPU clock speeds in the \emph{megahertz} range) compared to microprocessors.

\textbf{A flurry of multicore microcontrollers ---} Exploiting multicore efficiently is required to allow the execution of computation-intensive tasks, such as real-time audio processing or edge machine learning, 
while the device remains available for sensing, actuation, or push/pull of data over the network. 
Thus, many vendors' flagships are now based on multicore 32-bit architectures.  Examples include
the Espressif ESP32\nobreakdash-S3 microcontrollers based on a dual-core Xtensa LX7; the RP2350 microcontrollers based on dual-core RISC-V Hazard3 and dual-core ARM Cortex\nobreakdash-M33, used on popular boards such as the Raspberry Pi Pico 2. An earlier model, the RP2040, based on dual-core ARM Cortex\nobreakdash-M0+, had already sold millions of units. 
Other vendors  (Nordic, NXP, ST, etc.) also rolled out 32-bit multicore microcontrollers. 


\textbf{Embedded operating systems for microcontrollers ---} As software running MCUs and sensors became more complex, the use of an operating system (OS) became frequent. The most prominent operating systems are written in C (RIOT, Zephyr, FreeRTOS~\cite{hahm2015operating}).
Recently, a new breed of OS and embedded software platforms has emerged, written in Rust. 

This effort was pioneered by Tock~OS~\cite{levy2015ownership} and many \emph{bare-metal} embedded Rust programming efforts, which resulted in vastly improved embedded Rust~\cite{embedded-rust-book}. 
Examples of embedded Rust platforms for MCUs
include Hubris~\cite{hubris}, Drone~OS~\cite{drone-os}, RTIC~\cite{rtic}, as well as  asynchronous Rust with Embassy~\cite{embassy}. 
However, to date, none of these support multicore scheduling on the main 32-bit microcontrollers.

Aiming to bridge this gap:
\begin{itemize}
    \item We design \OSname{}, a novel embedded Rust operating system combining
    \begin{enumerate*}[label=(\roman*)]
        \item a scheduler exploiting single- and multicore  microcontrollers, and 
        \item asynchronous Rust;
    \end{enumerate*}
    \item We implement \OSname{} and provide benchmarks on diverse 32-bit microcontroller architectures: ARM \mbox{Cortex\nobreakdash-M}, Espressif Xtensa, and RISC-V;
    \item We overview the OS, which, beyond the scheduler, integrates diverse libraries and cross-hardware APIs;
    \item We publish the full \OSname{} code as open source. 
    
\end{itemize}

\section{Prior work on Multicore Scheduling on MCUs} \label{sec:multicore-sched}

Surveying existing software and prior literature, we found that notable examples of multicore scheduling on microcontrollers include ThreadX~\cite{threadx} and FreeRTOS~\cite{freertos}, Zephyr~\cite{zephyr} and NuttX~\cite{nuttx}.
Typically, threads that are ready and waiting for execution are listed  in a sorted data structure called the \emph{runqueue}. 
We observe two main approaches for runqueues on multicore architectures. The first approach, \textit{global scheduling}, uses a single global runqueue from which the threads are distributed onto the available cores.
In contrast, the \textit{partitioned scheduling} approach employs separate runqueues for each processor, to which threads are statically allocated.
Furthermore, we observed that OSs for microcontrollers that support multicore scheduling primarily use global scheduling.
We also observed that for synchronization within the scheduler, global critical sections are typically used.

We finally observed that operating systems in this space employ different approaches for assigning threads to cores. 
On the one hand ThreadX uses a \textit{core reallocation approach}, whereby a reallocation routine maps the \textit{n} highest priority threads to the \textit{N} cores. 
The allocation for each core is read by the scheduler interrupt handler upon invocation. 
After each change in the runqueue, the reallocation routine---in ThreadX called \textit{rebalance}---is executed.

On the other hand, FreeRTOS, Zephyr or NuttX use variants of a \textit{dynamic thread selection approach}, whereby the next thread for a core is selected directly from the runqueue by the scheduler interrupt handler upon invocation.
The thread is then removed from the runqueue to prevent it from being selected for multiple cores. Conversely, when a running thread is preempted, it is re-added to the runqueue. 
\section{Goals of \OSname{} and its Scheduler}
\label{sec:goals}

\textit{Use-Case \& Assumptions ---} Our work targets typical microcontrollers, which entails $N$ cores and $n$ threads, and only shared caches, if any.  Furthermore $N$ and $n$ are small: typically $N<3$ and $n<15$. Notice major differences versus Linux, L4 or HPC use-cases, which entails much larger $N$ and $n$, more elaborate cache and fairness mechanisms.



In this paper, we aim to 
support multicore, for Symmetric Multiprocessing (SMP) cases, which is the most common on microcontrollers so far, whereby the multiple cores are identical in their architecture and share the entire address space. 
We pursue the following objectives:
\begin{itemize}
    \item \textbf{Work-Conservation ---} 
    the scheduler must execute the \textit{n} highest priority threads on the \textit{N} available cores;
    \item \textbf{Real-Time ---} Retain the real-time properties of the scheduler, based on thread priorities set by the user;
    \item \textbf{Portability ---} Ensure portability between heterogeneous architectures and platforms.
    \item \textbf{Transparency ---} Transitioning from single-core to multicore use must be transparent, retain scheduler performance, and not increase complexity for the user.
\end{itemize}


\textit{Operating System ---} Beyond scheduling, \OSname{} aims to be a one-stop-shop for distributed computing and networked applications on heterogeneous 32-bit MCUs. More details on this bigger picture are given in Section~\ref{sec:user-guide}. 

\section{Single-Core Scheduler Basis \& Optimization}
\label{sec:single-core}

The \OSname{} builds upon a single-core scheduler basis similar to RIOT~\cite{baccelli2018riot}.
That is: a tickless real-time scheduler with a preemptive priority scheduling policy. 
The maximum number of threads is set at compile time, and all scheduler data structures are allocated statically.
Context switching is done lazily, in that a context is only switched when the head of the runqueue has changed since the last scheduler invocation.
The scheduler is further optimized to avoid superfluous scheduler interrupts.
Specifically, \OSname{} checks whether a new thread is ready and has higher priority than the current one \emph{before} setting a scheduler interrupt.

\section{\OSname{} Multicore Scheduler Design}
\label{sec:design-overview}
In the following, without loss of generality, we assume $N=2$ cores (Core 0 and Core 1). Note, however, that scaling to $N>2$ is trivial, assuming $N$ and $n$ remain small as per our goals (recall~\autoref{sec:goals}).

\subsection{Startup}

\begin{figure}
\centering
\includegraphics[width=0.95\columnwidth]{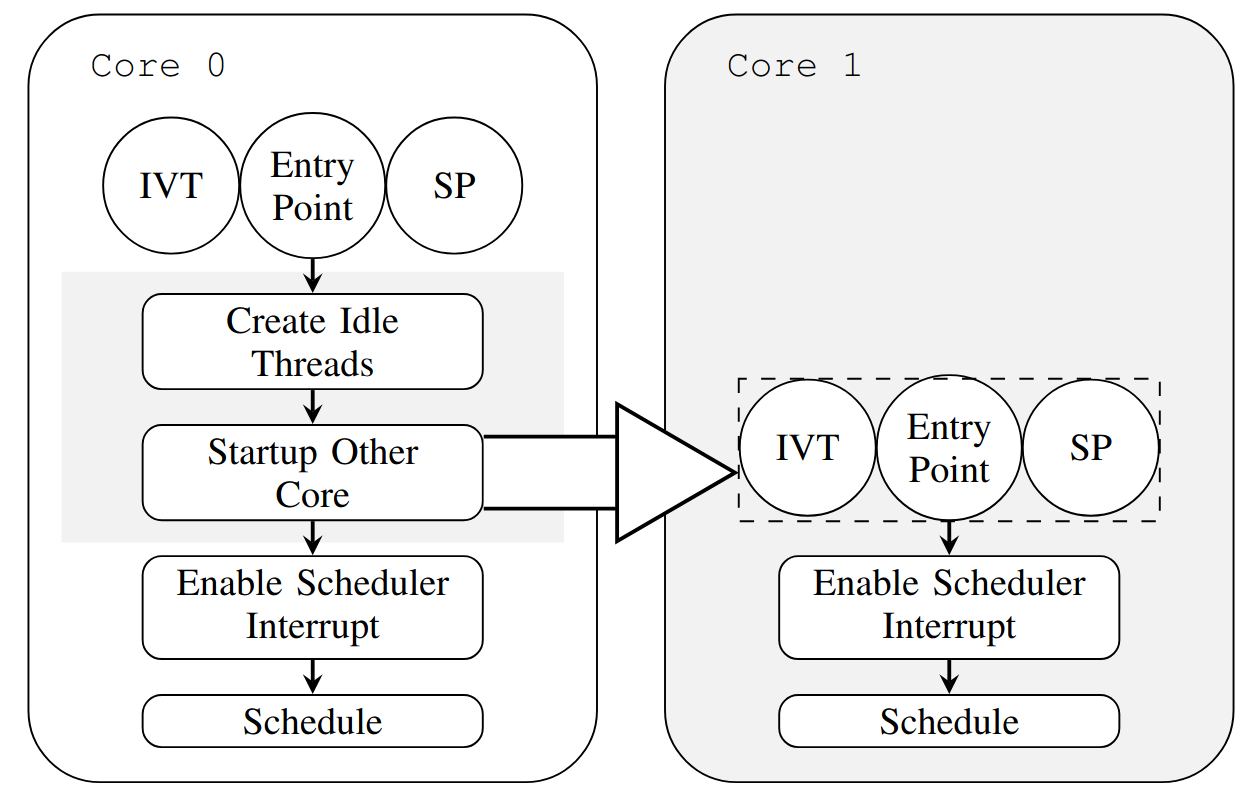}
     \caption{Threading startup on a dual-core system.}
     \label{fig:startup}
 \end{figure}

The \OSname{} runtime initialization executes on Core 0, while the other core is in deep sleep or stall mode. 
After the general initialization on Core 0, Core 1 is set up with the shared interrupt vector table (IVT), a separate main stack pointer (SP), and an entry function that, down the line, will start threading on this core, analogous to Core 0.
This is depicted in Fig.~\ref{fig:startup}.
Threading is started by enabling the scheduler interrupt at the lowest priority and invoking the scheduler, which will then select the next thread for the core.



\subsection{Global Scheduling Scheme}\label{sec:design:scheduling}


\OSname{} uses a global scheduling scheme.
Global scheduling facilitates thread migration, where threads migrate between cores between executions.
We chose global scheduling because it reduces context switches and priority inversions, and makes better use of the overall capacity~\cite{hardrealtime-multiprocessor-survey, brandenburg}.
Furthermore, it avoids the thread allocation problem of partitioned scheduling, i.e., the problem of finding an optimal allocation of threads to cores.
The main drawback of global scheduling is scalability, when the global runqueue becomes a bottleneck for context switching~\cite{EDF-RM-multiprocessor-survey}. However, this issue is negligible on MCUs --- which have a low processor count.

\subsection{Assigning Threads to Cores}
\label{sec:Assigning-Threads-to-Cores}
With the single-core configuration, the thread-selection process simply reads the head of the runqueue. 
However, on multicore, this would result in the same thread being executed on multiple cores.
There are various approaches for selecting the next thread that a core should execute.
We initially designed our scheduler so that it can accommodate either \textit{core reallocation} or alternatively \textit{dynamic thread selection} (see \autoref{sec:multicore-sched}). We then implemented both, before reporting on their performance on different hardware in \autoref{sec:benchmarks}.
\subsection{Core Affinity Masks}

Core affinities are a feature that allows pinning a thread so it runs on some specified core(s) only. 
This provides the user with fine-grained control over how and where threads may run. 
Core affinities are realized with a bitmask, where a set bit encodes that the thread can run on the core with this index. 
The scheduler uses this information when selecting the core on which a thread should run.

\subsection{Dynamic Priorities and Priority Inheritance}

Thread priorities in \OSname{} can be changed dynamically at runtime for all active threads.
A change in priority may result in a context switch if a running thread's priority is decreased or if a waiting thread's one increased.
Furthermore, dynamic priorities facilitate priority inheritance for mutexes.
If a higher-priority thread is blocked on a mutex that is currently owned by a lower-priority thread, the owning thread inherits the higher priority of the waiting thread.
This prevents the owner from being preempted by other lower-priority threads 
and thus helps avoid priority inversions caused by indirect blocking.

\subsection{Synchronization in the Scheduler\label{sec:design:sync}}

\OSname{} implements mutual exclusion in the kernel through a global critical section, effectively resulting in a \textit{Big Lock} design.
Thus, there is no concurrency in the kernel, which is tolerable given that such operations in \OSname{} are short.
It ensures correctness, prevents data races and deadlocks, and simplifies operations that involve multiple data structures, such as the runqueue and Thread Control Blocks (TCBs), because no individual locking is needed.
On single-core systems, it is sufficient to mask all interrupts to create a critical section.
In multicore systems, an additional global spinlock is required to ensure that even threads on other cores cannot enter another critical section. 

\subsection{Combining Scheduled Threads \& Async Rust Tasks}\label{sec:design:async}

\OSname{} uses async code based on \emph{Embassy}~\cite{embassy} for system initialization and in the HAL.  
There is always a system executor for async tasks. 
Using the preemptive scheduler --- and thus threading --- is optional. 
If the scheduler is not used, the system executor will run in interrupt context. 
If the scheduler is used, the system executor executes in a thread. 
Additional executors can be started in other threads. 
When all tasks on an executor are pending, the owning thread is suspended. 
Threads can block on async functions and wait for async resources from an executor.
Thus, the gap is bridged between the scheduler, async Rust, future-based concurrency, and asynchronous I/O.

%


\section{\OSname{} Scheduler Implementation}


As of this writing, \OSname{} supports multicore scheduling on three popular hardware platforms/architectures: the RP2040 and RP2350 (dual-core ARM Cortex-M0+/ dual-core ARM Cortex-M33 resp.) and the \espsthree{} (dual-core Xtensa LX7). RISC-V multicore scheduling support in \OSname{} is currently a work in progress. 

The RP2040, RP2350, and the \espsthree{} are well supported in the Rust ecosystem: the former through the Embedded-Rust working group and the \verb|rp-rs| project, the latter directly by Espressif through the \verb|esp-hal| project.
\OSname{} leverages this support in its implementation.

\subsection{Hardware Abstraction}


One of the main objectives of our work is clear hardware abstraction and the reduction of platform-specific code.
Adding support for another chip should be minimal effort, particularly when there is already chip support in the ecosystem.
In \OSname{}, hardware abstraction occurs at two levels:

\textit{CPU architecture abstraction ---} At this layer, the scheduler logic for a CPU architecture, e.g., Cortex-M, is implemented. 
It involves all architecture-specific code to set up a thread stack, configure the exception used to trigger the scheduler, and the actual context-switching logic. 
    
\textit{Chip-level abstraction ---} The platform-specific logic for SMP is implemented at this layer.
    For multicore scheduling, two mechanisms are required,
    \begin{enumerate*}[label=(\roman*)]
    \item for starting up the other core(s), and 
    \item for invoking the scheduler on a specific core
    \end{enumerate*}. 

\OSname{} takes advantage of the Rust type system, by specifying the above abstractions as \emph{traits}~\cite{rust-book_traits}.
Adding support for another platform in the scheduler only requires two traits -- one per layer -- to be implemented.
As a result, the corresponding chip-level SMP implementations in \OSname{} are only 70 lines of Code (LOC) for the RP2040 and 66~LOC for the \espsthree{}.

\subsection{Platform-specific Scheduling Logic}

\noindent \textit{On the RP2040/RP2350 ---} The chip implements (in hardware) two FIFO queues between the physical cores, which are utilized by \OSname{} during startup and for scheduler invocations.
During startup, Core 1 remains in sleep mode until it receives the vector table, stack pointer, and entry function through the FIFO queue, based on a fixed protocol.
After startup, the FIFO queue is used to invoke the scheduler on the other core. 
A received message will trigger an interrupt, where the handler will set the local scheduler exception.

\noindent \textit{On the ESP32-S3 ---} The chip does not implement any inter-processor communication channel.
Instead, two different software-triggered CPU interrupts are used to invoke a scheduler on each core, respectively.
Startup of the second core on the \espsthree{} is implemented by writing the address of the entry function into the boot address of Core 1, then resetting and unstalling the core.
The entry function sets up the vector table address and the stack pointer for this core, and then runs our threading startup logic.

\subsection{Interrupt Handling}

On the RP2040/RP2350, each core is equipped with its own ARM Nested Vectored Interrupt
Controller (NVIC). 
On the \espsthree{}, each core has its own configurable interrupt matrix. 
On both chips, nested interrupts are supported, so that a lower-priority interrupt can be preempted by a higher-priority one.
External interrupts are routed to both cores,
but only enabled on one core for handling, as described below.

During startup, peripherals are initialized either in interrupt mode on Core 0 before threading is started, or by one high-priority thread. 
This initialization configures and enables the required interrupts on the core it executes on, which will then be the core that handles the related interrupts.

Still, both cores share the same interrupt vector table, so it is also possible to manually mask and unmask interrupts on individual cores to configure where an interrupt should be handled.

\subsection{Synchronization in the Scheduler}

The scheduler is implemented through a single structure that contains the runqueue, TCBs, and other scheduling data, and implements all scheduling logic.
This structure is protected by a wrapper type that ensures that all accesses to the scheduler are executed inside a critical section and that no reference to the scheduler can be obtained outside of it.
\section{Micro-Benchmarks with \OSname{} Multicore}
\label{sec:benchmarks}

We next evaluate the performance of \OSname{} on popular hardware. We published our benchmark code in~\cite{ariel-benchmarks}. 

The \textbf{dual-core MCUs} we used are:
\begin{enumerate*}[label=(\roman*)]
\item Espressif \espsthree{} with dual Xtensa LX7 at 240 MHz, and 
\item RP2040/RP2350 with dual Cortex-M0+/M33 at 133/150 MHz.
\end{enumerate*}

The \textbf{single-core boards} we used for comparison are:
\begin{enumerate*}[label=(\roman*)]
\item a Nordic nRF52840 with a Cortex-M4 at 64 MHz, and 
\item an Espressif \espcthree{} with a RISC-V RV32IMC at 160 MHz.
\end{enumerate*}

On Xtensa and Cortex-M, the performance is measured in ticks.
On the RISC-V-based \espcthree{}, we instead use the system timer, which runs at 16 MHz.
The measurements we present are the average of 1000 runs.

\subsection{Comparing Multicore Scheduler Variants}

\autoref{fig:flags} depicts micro-benchmarks in which four threads, grouped into pairs, alternate in waking each other up and then suspending their own execution. 
We compare the performance of different variants of the scheduler: (i) the \emph{single-core} variant described in \autoref{sec:single-core}, (ii) a multicore variant using \emph{core reallocation}, and (iii) a multicore variant using \emph{dynamic thread selection} as mentioned in \autoref{sec:Assigning-Threads-to-Cores}. 

\pgfplotstableread[col sep=comma,]{data/flags_dual-core_espressif-esp32-s3-devkitc-1.csv}\flagsdualesp{}
\pgfplotstableread[col sep=comma,]{data/flags_dual-core_rpi-pico.csv}\flagsdualpico{}
\pgfplotstableread[col sep=comma,]{data/flags_dual-core_rpi-pico2.csv}\flagsdualpicotwo{}
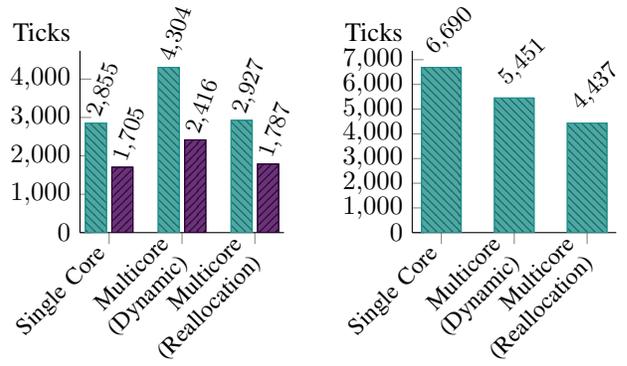
\begin{figure}[t!]
     \begin{subfigure}[t]{0.49\columnwidth}
     \begin{tikzpicture}
         {
            \begin{axis}[
            ybar,
            ourybarstyle,
            bar width=8pt,
            enlarge x limits=0.2,
            height=4cm, width=\textwidth,
            ylabel={Ticks},
            ytick distance=1000,
            symbolic x coords={{Single Core},{Multicore (Reallocation)},{Multicore (Dynamic)}},
            xticklabels={{Single Core},{Multicore\\(Reallocation)},{Multicore\\(Dynamic)}},
            nodes near coords style={font=\small, black, rotate=70, anchor=south east, yshift=-6pt, xshift=+27pt},
            xticklabel style={font=\small, rotate=45, anchor=east, align=right},
        ]
      \addplot table [y index=1]{\flagsdualpico};
      \addplot table [y index=1]{\flagsdualpicotwo};
    \end{axis}}
    \end{tikzpicture}
    \caption{RP2040/RP2350 (blue/purple)\label{fig:flags_rp2040}}
     \end{subfigure}
     \begin{subfigure}[t]{0.49\columnwidth}
     \begin{tikzpicture}
         {
            \begin{axis}[
            ybar,
            ourybarstyle,
            bar width=15pt,
            enlarge x limits=0.2,
            height=4cm, width=\textwidth,
            ylabel={Ticks},
            ytick distance=1000,
            symbolic x coords= {{Single Core},{Multicore (Reallocation)},{Multicore (Dynamic)}},
            xticklabels={{Single Core},{Multicore\\(Reallocation)},{Multicore\\(Dynamic)}},
            nodes near coords style={font=\small, black, rotate=45, anchor=south west, xshift=-2pt},
            xticklabel style={font=\small, rotate=45, anchor=east, align=right},
        ]
      \addplot table [y index=1]{\flagsdualesp};
    \end{axis}}
    \end{tikzpicture}
    \caption{ESP32-S3 (dual Xtensa LX7)\label{fig:flags_esp32s3}}
    \end{subfigure}
    \caption{Context switching performance of the two multicore scheduler designs compared with single-core configuration.\label{fig:flags}}
\end{figure}

The results show that \textit{dynamic} outperforms \textit{reallocation} on all the multicore microcontrollers we tested.
The relatively more complex \textit{reallocation} routine, executed after each thread state change, takes a toll. 

The benchmark largely consists of scheduler operations. Thus, mutual exclusion in the scheduler forces mostly sequential execution on the RP2040.
In contrast, on the \espsthree{}, parallelization happens differently in hardware, forcing sequential execution less often. 
Furthermore, one benchmark iteration on \espsthree{} requires
many more ticks than on the RP2040.
Thus, \espsthree{} reaps more benefits from parallelization in this benchmark.
In the following, we focus on the \textit{dynamic thread selection} variant for our multicore scheduling.

\subsection{Overhead of Multicore Scheduling}

We next measure
the overhead of our multicore scheduling feature, compared to the single-core scheduler we started from in \autoref{sec:single-core}.
For this, we focus on thread preemption, because, compared to single-core, the multicore feature adds an additional step: inserting the preempted thread back into the runqueue. 
\autoref{fig:preempt} reports on micro-benchmarks performed on different single-core boards, in which a lower-priority thread sets the flag for a higher-priority thread, resulting in a context switch.
We observe that the overhead when the multicore feature is enabled remains quite small, approx. 9.6\% on the nRF52840 and 5.3\% on the \espcthree{}. 
Conveniently, we can therefore enable multicore by default
in \OSname{}.

\pgfplotstableread[col sep=comma,]{data/preempt_both_ai-c3.csv}\preemptesp{}
\pgfplotstableread[col sep=comma,]{data/preempt_both_nrf52840dk.csv}\preemptnrf{}
\begin{figure}[t!]
     \begin{subfigure}[t]{0.49\columnwidth}
     \begin{tikzpicture}
         {
            \begin{axis}[
            ybar,
            ourybarstyle,
            bar width=20pt,
            enlarge x limits=0.4,
            height=4cm, width=\textwidth,
            ylabel={Ticks},
            ytick distance=200,
            symbolic x coords={{Multicore scheduler disabled},{Multicore scheduler enabled}},
            xticklabels={{disabled},{enabled}},
            nodes near coords style={font=\small, black, rotate=45, anchor=south west, xshift=-2pt},
            xticklabel style={font=\small, align=center},
        ]
      \addplot table [y index=1]{\preemptnrf};
    \end{axis}}
    \end{tikzpicture}
         \caption{nRF52840 (Cortex-M4)\label{fig:preempt_nrf52840}}
     \end{subfigure}
     \begin{subfigure}[t]{0.49\columnwidth}
     \begin{tikzpicture}
         {
            \begin{axis}[
            ybar,
            ourybarstyle,
            bar width=20pt,
            enlarge x limits=0.4,
            height=4cm, width=\textwidth,
            ylabel={Ticks},
            ytick distance=50,
            symbolic x coords={{Multicore scheduler disabled},{Multicore scheduler enabled}},
            xticklabels={{disabled},{enabled}},
            nodes near coords style={font=\small, black, rotate=45, anchor=south west, xshift=-2pt},
            xticklabel style={font=\small, align=center},
        ]
      \addplot table [y index=1]{\preemptesp};
    \end{axis}}
    \end{tikzpicture}
         \caption{ESP32-C3 (RISC-V)\label{fig:preempt_esp32c3}}
    \end{subfigure}
    \caption{Overhead of the multicore scheduling feature, measured on single-core hardware. 
    \label{fig:preempt}}
\end{figure}

\subsection{Speedup of Computation Intensive Tasks}

\pgfplotstableread[col sep=comma,]{data/matrix-mult_speedup_espressif-esp32-s3-devkitc-1.csv}\matrixesp{}
\pgfplotstableread[col sep=comma,]{data/matrix-mult_speedup_rpi-pico.csv}\matrixpico{}
\pgfplotstableread[col sep=comma,]{data/matrix-mult_speedup_rpi-pico2.csv}\matrixpicotwo{}
\begin{figure}[t]  
     \begin{subfigure}{0.49\columnwidth}
     \begin{tikzpicture}                                                   
         {
            \begin{axis}[                  
            ourlinestyle,              
            height=4cm, width=\textwidth,  
            ylabel={Relative speedup},     
            ytick distance=0.5,         
            tension=0.1,                
        ]                               
        \addplot+[myTeal, mark options={fill=myTeal}] table [x=0, y=speedup] {\matrixpico};                    
        \addplot+[myEggplant, mark options={fill=myEggplant}] table [x=0, y=speedup] {\matrixpicotwo};
    \end{axis}}                           
    \end{tikzpicture}      
     \caption{RP2040/RP2350 (blue/purple)\label{fig:matrix-mult_rp2040}}
     \end{subfigure}                                                   
     \begin{subfigure}{0.49\columnwidth}                                                                            
     \begin{tikzpicture}                                                                                                          
         {                                                                                                              
            \begin{axis}[                                                                                    
            ourlinestyle,                                                                             
            height=4cm, width=\textwidth,
            ylabel={Relative speedup},
            ytick distance=0.5,
        ]
        \addplot+[myTeal, mark options={fill=myTeal}] table [x=0, y=speedup] {\matrixesp};
    \end{axis}}             
    \end{tikzpicture}
         \caption{ESP32-S3 (dual Xtensa LX7)\label{fig:matrix-mult_esp32s3}}
    \end{subfigure}                                                             
    \caption{Multiplication of \(N\times N\) matrices, \(N\in\{10, 20, ...,  80\}\)\label{fig:matrix-mult}}
\end{figure}  
We next benchmark 
$N\times N$ matrix multiplication, with and without the multicore feature. 
The computation is split into 2 tasks, where the 1st and 2nd halves of the first matrix are each multiplied with the second matrix.
In the single-core configuration, the two computations are executed sequentially.
For multicore, the tasks are distributed to two threads that are scheduled in parallel.
We observe in Fig.~\ref{fig:matrix-mult} that when $N$ is small, IPC overhead dominates parallelization gains. When $N$ grows larger, we observe a performance gain tending towards 2x.
We also notice that performance gains are not strictly linear. This is subject for further investigation --- it might be due to MCU-specific memory subsystem characteristics. 
For example, on the RP2040, both cores compete for a shared memory bus~\cite{pico-mem-layout}.

\textbf{Summary ---} By using a global scheduling scheme and priorities, we achieve \emph{work-conservation} and \emph{real-time} properties. Our hardware abstraction provides \emph{portability}, and we demonstrated \emph{transparency} via our experimental measurements. We have thus achieved the goals we set in Section~\ref{sec:goals}. 

\section{Ariel Operating System Overview 
}\label{sec:user-guide}

 \begin{figure}[t!]
     \centering
     \includegraphics[width=0.7\linewidth]{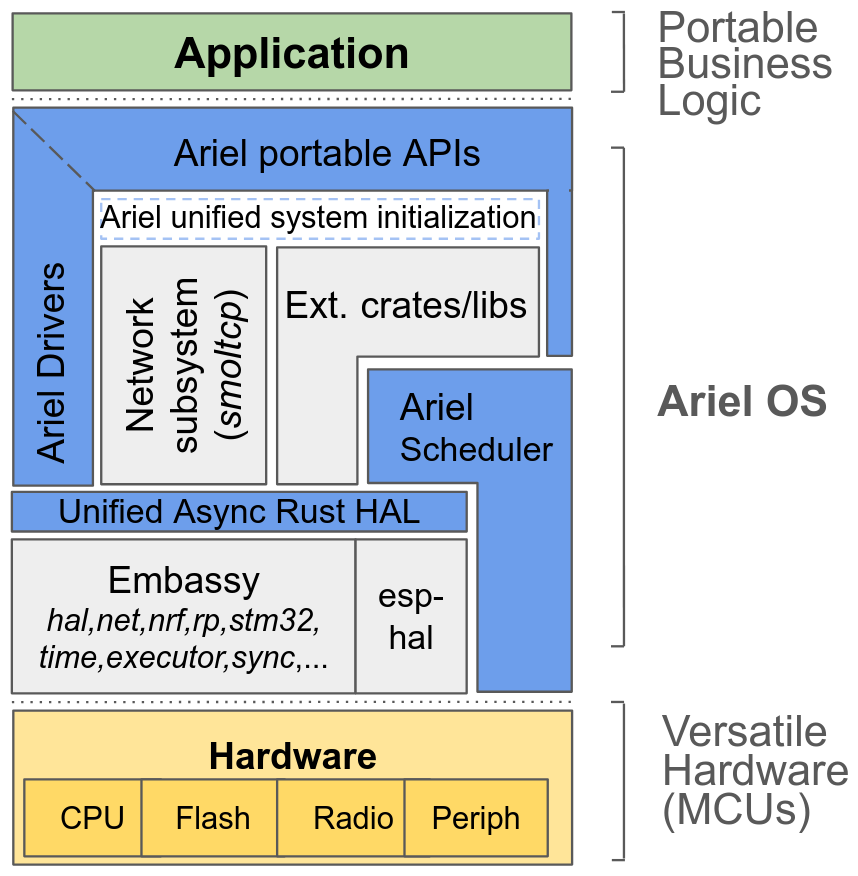}
     \caption{\OSname{} architecture diagram.
     }
     \label{fig:ariel-arch}
 \end{figure}

\OSname{} is fully open-source~\cite{ariel-os-repo}. For basic hardware abstraction and async Rust programming, \OSname{} builds on top of Embassy~\cite{embassy}. Fig. \ref{fig:ariel-arch} shows how \OSname{} components (in blue) harness the ecosystem of embedded Rust (in grey).
In particular, \OSname{} combines the following elements:

\textbf{Versatile network stack configurations ---}
\OSname{} integrates a network stack (\emph{embassy-net/smoltcp}~\cite{smoltcp}) combined with additional modules we provide allowing various network configurations. These configuration options include IPv4 and IPv6, HTTP/TCP and CoAP/UDP, over wireless and wired link layers, secured by open standards including COSE, OSCORE, EDHOC, TLS~\cite{tschofenig2019cyberphysical} and a curated set of libraries providing cryptographic backends.

\textbf{Abstracted system initialization ---}
On MCUs, code initializing the system can be very challenging for developers. \OSname{} thus abstracts initialization, e.g., setting up
\begin{enumerate*}[label=(\roman*)]
\item the network stack, 
\item cryptographic material and identities, 
\item the random number generator, 
\item USB peripherals,
\end{enumerate*}
etc.
Configuration is handled at build system level. Convenient defaults are provisioned. Boilerplate is thus minimized and high-level building blocks are provided to application logic. 

\textbf{Unified peripheral APIs ---} 
\OSname{} crafted peripheral initialization and setup (for GPIO, I2C, SPI accessing sensors/actuators) to be identical across MCU families. Thereby, application code written once can be compiled for all devices that \OSname{} supports. 
This is a substantial improvement over the state of the art in embedded Rust (crates such as \emph{embedded-hal} or \emph{embedded-hal-async}) which leave out initialization, and thus lead to a jungle of initialization APIs --- and very limited application code portability. 

\textbf{Meta build system ---}
The \OSname{} toolchain takes full advantage 
of the embedded Rust \emph{crates} ecosystem and the Rust build system \emph{Cargo}. To work around limitations w.r.t. extremely diverse modular target configurations, we wrapped Cargo in \emph{laze}~\cite{laze}, our meta-build system handling the huge matrix of software configurations on various boards.

\section{Conclusions}
In this paper we introduced \OSname{}, the first embedded Rust operating system for microcontrollers supporting both single- and multicore preemptive scheduling combined with asynchronous Rust.
We assessed experimentally how 
a unique multicore scheduler can be the convenient default on all supported multicore platforms. 
Still, applications 
can opt out of multicore scheduling when parallelization is unneeded.
\OSname{} thus enriches the set of available open source tools 
for secure and efficient distributed computing applications involving sensors/actuators or small networked devices using 32-bit MCUs such as ARM Cortex\nobreakdash-M, RISC\nobreakdash-V, or ESP32.





\bibliographystyle{IEEEtran}
\bibliography{bibliography}



\vspace{12pt}

\end{document}